\documentclass[12pt]{article}
\usepackage{amsmath,amssymb}
\usepackage[dvips]{graphicx}

\begin{document}

\begin{center}

\section*{The   turbulence development  at {its} initial stage: a scenario  based on the {idea} of vortices decay.}

{S.V. TALALOV}

{Department of Applied Mathematics, Togliatti State University, \\ 14 Belorusskaya str.,
 Tolyatti, Samara region, 445020 Russia.\\
svt\_19@mail.ru}

\end{center}


\begin{abstract}
In this paper, a model of the development of a quantum turbulence in its  initial stage is proposed.
The origin of the turbulence in the suggested model is the decay of vortex loops with an internal structure.
We consider the initial stage of this process, before  an  equilibrium state is established.
As result of  our study, the density matrix of developing turbulent flow is calculated.
 The quantization scheme of the  classical vortex rings system is based on the approach proposed by the author earlier. 
\end{abstract}

{\bf keywords:}   quantum vortex rings,  quantum vortex interaction, quantum fluids  

\vspace{5mm}


\section{Introduction}

~~~The description of turbulence at the quantum level has been a complicated issue up till now. 
{ As it was noted in the paper \cite{PoMuKr_1}, ''\dots the understanding of turbulent flows is one of the biggest current challenges in physics, as no first-principles theory exists to explain their observed spatio-temporal intermittency''. A numerical simulation of turbulence remains one of the main challenges in the field of study 
  of this complicated phenomenon. A large number of papers is devoted to it.  Many of these papers use the Gross-Pitaevskii model as the base for constructing the corresponding algorithms. (see, for example, \cite{GiuKrs}).  Modern methods are also used: in the paper  \cite{Fuka}  quantum computing algorithms were  developed for the turbulence simulation purposes.
One of the important properties of the turbulence flow  is an extremely
wide  range of length and time scales. This question was discussed, for example, in the paper
 \cite{Gourianov}.  The difference between classical and quantum turbulence is discussed in the work \cite{madeira}. This question has a long history. As it was emphasized in the review
\cite{Vinen}, the differencies between the classical and quantum turbulence are due to the quantum effects. }
Within the framework of this topic, there are many directions  for the research
 \cite{PrLTPh}.
It is now an established fact that vortex structures play a primary role in the formation of turbulent flows in quantum fluids.
A large number of works is devoted to this issue \cite{Feyn,Donn,TsFuYu,Nemir,MuPoKr}. 
Out of all topics covered in the literature, the process of the formation of a turbulent flow itself remains the least studies one.
In this paper we consider a quantum model which describes  the  emergence of a turbulent flow in its initial stage.
The basic  assumption of our study
 is  a quite standard one:  the reason for the turbulence appearance is the series of the vortexes decays.  In this  paper, we are going to provide  a model that describes the mechanism of such processes.
The  results that will be obtained here are a natural development of the new approach which has been elaborated by the author\cite{Tal}.
  This approach contains  both quantum description of a single vortex loop and 
the quantum description of a many-vortex system.

\section{Initial assumptions}

In this section we will formulate the initial assumptions for constructing the  subsequent theory. Moreover, we will give a brief overview of the author's approach to quantization of vortex rings.
 Details of the approach can be found in the papers \cite{Tal,Tal_PoF}. 
We will use the following dimensional constants in our theory: the fluid's density $\varrho_0$, the speed of sound in this fluid $v_0$ and the natural scale length $R_0$:
$$R_0 \in  \{\,R\,:   R =   |\,{\boldsymbol r}_1 - {\boldsymbol r}_2|\,, \qquad {\boldsymbol r}_1, {\boldsymbol r}_2 \in V\, \}\,,$$
 where   symbol  $V$  denotes the domain where the investigated objects evolve.
If we consider an ideal case of   motion in an unbounded space, the constant $R_0$  must be
defined   in some additional way.
The  value  $\tilde\mu_0 =  \varrho_0 R_0^3$   is a natural parameter that determines the scale of the masses in the considered model. Despite this fact,
  we will  also   use  the  additional  mass   parameter $\mu_0 $ here.
	In our theory this parameter denotes
 the central charge for central extension of the Galilei group  ${\mathcal G}_3$ ({The appearance of extended Galilei group in the considered approach was discussed in the author's paper \cite{Tal} in detail. }). Thus, the additional dimensionless parameter    $\alpha_{\sf ph} = \mu_0/\tilde\mu_0$ appears.
		Such a parameter will provide our theory with additional possibilities in the subsequent descriptions of possible observed effects.  	
For convenience, we will use the auxiliary  constants $t_0 = R_0/v_0$ and   ${\cal E}_0  = \mu_0 v_0^2$ along with the constants $\varrho_0$, $v_0$, $R_0$.

On a classical level, we consider  the special configurations of the closed vortex filament $ {\boldsymbol{r}}(t ,s)$ with an internal  core structure. 
The core  radius   ${\sf a}$ is assumed to be finite, but small enough to use the   ''vortex filament''  approximation.
We suppose that the dynamics of  such objects  can be described  by the local induction equation.    
Before writing down this equation,
we define  the projective vectors ${\boldsymbol{r}}   \to {\boldsymbol{r}}/R$, where symbol $R$ denotes the arbitrary positive constant with a dimension of length. 
 For convenience,  we introduce the  dimensionless parameters
$\tau = t/t_0$ and $\xi = s/R$  to describe the considered vortex filament.  Finally, the dynamical equation for the vortex filament is written as follows

\begin{eqnarray}
        \label{LIE_str}
      ~&~&~  \partial_\tau {\boldsymbol{r}}(\tau ,\xi)   = 
        \beta_1 \Bigl(\partial_\xi{\boldsymbol{r}}(\tau ,\xi)\times\partial_\xi^{\,2}{\boldsymbol{r}}(\tau ,\xi)\Bigr)   + \nonumber \\
				~~ & + & \beta_2\Bigl(2\,\partial_\xi^{\,3}{\boldsymbol{r}}(\tau ,\xi) + 
        {3}\,\bigl\vert\, \partial_\xi^{\,2}{\boldsymbol{r}}(\tau ,\xi)\bigr\vert^{\,2}\partial_\xi{\boldsymbol{r}}(\tau ,\xi)\Bigr)\,.
				        \end{eqnarray}
			
		This equation simulates the dynamics of a vortex filament  with an internal flow in the core (for the case $\beta_2 \not= 0$). It was investigated in the Ref. \cite{AlKuOk}.						
The  values  $\beta_1$ and $\beta_2$  are dimensionless constants here. 
Eq. (\ref{LIE_str}) has  { definite}    
{ scale - invariant}
solution that is of interest to the proposed model. This solution is:

\begin{equation}
        \label{our_sol}
 \boldsymbol{r}(\tau ,\xi) = \Bigl(\, \frac{q_x}{R} +  \cos(\xi +\phi_0 +\beta_2\tau)\,, \frac{q_y}{R} +  \sin(\xi +\phi_0+\beta_2\tau  )\,, \frac{q_z}{R} + \beta_1 \tau \,\Bigr)\,. 
\end{equation}

where the angle   $\phi_0 \in [0, 2\pi)$ and the coordinates $q_x, q_y, q_z$   are some
 (time - independent) variables. 
This solution describes the vortex filament  in the shape of a circle with a radius $R$.   The filament  moves along the axis  $\boldsymbol{e} = {\boldsymbol{e}}_z$
with velocity $|\boldsymbol{u}_{v}| = \beta_1 R/t_0$ and  rotates with  the frequency  $\beta_2/t_0$.  This rotation  simulates  certain  flow $\Phi$  inside the  filament  core.
This flow is:
\begin{equation}
        \label{Flow_class}
				\Phi =  \beta_2\pi \varrho_0{\sf a}^2 v_0  \bigl(  R/R_0\bigr)\,. \nonumber
\end{equation}
Further, we will only consider 
 such solutions for the Eq. (\ref{LIE_str}). Thus, the set of possible vortex loops is reduced  to the rings of an arbitrary radius and some fluid flow in the core.


In addition to the Eq. (\ref{LIE_str}) that describes the evolution of the  
 curve ${\boldsymbol{r}}(\cdot,\xi)$, we postulate the standard hydrodynamic formula \cite{Batche}
for the momentum   $\tilde{\boldsymbol{p}}$:
	  
   \begin{equation}
        \label{p_and_m}
        \tilde{\boldsymbol{p}} = \frac{\varrho_0}{2 }\,\int\,\boldsymbol{r}\times\boldsymbol{\omega}(\boldsymbol{r})\,dV\,.
                 \end{equation}
The vorticity $\boldsymbol{\omega}(\boldsymbol{r})$ has the following form for a vortex filament:
\begin{equation}
        \label{vort_w}
     \boldsymbol{\omega}(\boldsymbol{r}) =  \Gamma
                  \int\limits_{0}^{2\pi}\,\delta(\boldsymbol{r} - \boldsymbol{r}(\xi))\partial_\xi{\boldsymbol{r}}(\xi)d\xi\,,
       \end{equation}  
where the symbol 	$\Gamma$ denotes the circulation. 
The integral in r.h.s. of the  formula Eq. (\ref{p_and_m}) is easily calculated for solution Eq. (\ref{our_sol}). Simple calculations lead to the result
\begin{equation}
        \label{impuls_our1}
                \tilde{\boldsymbol{p}}  =   \pi \varrho_0 {R}^2 \Gamma  {\boldsymbol{e}} \,,\qquad   |{\boldsymbol{e}}|  = 1\,,                
                       \end{equation} 
where  constant unit vector  ${\boldsymbol{e}}$  defines the axis of the rotating ring Eq. (\ref{our_sol}). 

As can be seen from the formulas Eqs. (\ref{our_sol})  and   (\ref{p_and_m}), the natural variables that parametrize   our dynamical system are  variables
\begin{equation}
        \label{var_in}
 \boldsymbol{q} =  (\,q_x\,, q_y\,, q_z\,)\,,\quad R\,,\quad 
 \phi(\tau) = \phi_0 +\beta_2\tau \,,\quad \Gamma\,, \quad {\boldsymbol{e}}\,,  
\end{equation} 
where   $|{\boldsymbol{e}}|  = 1$. 
Here it is necessary to emphasize the following circumstance. 
Inclusion  of the value $\Gamma$   in the number of dynamic  variables     allows us to take into account the movement of the fluid surrounding the vortex  filament.
 Indeed, the consideration of the Eq. (\ref{LIE_str}), without  Eq. (\ref{p_and_m}),  describes some formal dynamics of the curve only.

However, the variables Eqs. (\ref{var_in}) are not well suited for the subsequent quantization of the theory.
 Let us define the variables
$$  \varpi = \frac{R}{R_0}\cos(\phi_0 +\beta_2\tau)\,, \qquad   \chi = \frac{R}{R_0}\sin(\phi_0 +\beta_2\tau)\,.$$
Dynamical equations for these variables are canonical Hamiltonian equations for a harmonic oscillator:
\begin{eqnarray}
\partial_\tau  \varpi & = &  -  \beta_2 \chi\,, \nonumber\\
  \partial_\tau  \chi  & = &   \beta_2 \varpi\,. \nonumber
\end{eqnarray}
As the following step, we introduce a vector  ${\boldsymbol{p}}   =  \alpha_{\sf ph}\tilde{\boldsymbol{p}}$ instead the ''canonical'' momentum  $\tilde{\boldsymbol{p}}$.
Consequently, we can rewrite the formula Eq. (\ref{impuls_our1})  as follows:
\begin{equation}
        \label{impuls_our2}
                 {\boldsymbol{p}}  =   \pi \alpha_{\sf ph} \varrho_0 {R_0}^2 \Gamma \bigl(\varpi^2  +  \chi^2\bigr){\boldsymbol{e}} \,,\qquad   |{\boldsymbol{e}}|  = 1\,.                
                       \end{equation}

From the author's point of view, the set of the variables ${\boldsymbol{p}}$,  $\boldsymbol{q}$, $\varpi$  and  $\chi$ adequately describes  our dynamical system Eq. (\ref{our_sol})
 as a structured  $3D$ particle with an internal degree of the freedom. 
The formula Eq. (\ref{impuls_our2}) together with the definition of the variables  $\varpi$  and  $\chi$ provides one-to-one correspondence  between the set Eq. (\ref{var_in})
and the  set (${\boldsymbol{p}}$,  $\boldsymbol{q}$, $\varpi$, $\chi$).  Note that the variables $\varpi$ and $\chi$  are invariants under Galilean and scale transformations of space $E_3$.

{The group of the space - time invariance in our theory is Galilei group  ${\mathcal G}_3$.  
In order to use a group-theoretical  approach to the definition of energy, we  consider a one-parameter central extension of such a group. 
 Indeed, Lee algebra $\widetilde{\mathcal G}_3$  of the  group ${\mathcal G}_3$ has following Cazimir functions:  
	
	 \[ {\hat C}_1 = \mu_0 {\hat I}\,,\qquad 
  {\hat C}_2 = \left({\hat M}_i  - \sum_{k,j=x,y,z}\epsilon_{ijk}{\hat P}_j {\hat B}_k\right)^2 \,,\]
	  \[ {\hat C}_3 = \hat H -  \frac{1}{2\mu_0}\sum_{i=x,y,z}{\hat P}_i^{\,2}\,,\]                   
where   the value  $\mu_0$ is the central charge and the values  ${\hat M}_i$,   $\hat H$,  ${\hat P}_i$         and  ${\hat B}_i$  ($i = x,y,z$)
        are the  generators of rotations, time and space translations and Galilean boosts respectively. 
		As usual, the function  ${\hat C}_3 $  can be interpreted as  an  ''internal energy of the particle''. 
		In our case it is naturall postulate that 		      			
       $${ C}_3  			=   {\beta_2{\cal E}_0} |\,b\,|^2\,, \qquad b = \frac{ \chi   + i \varpi}{\sqrt{2}}    	\,.$$ 
Therefore, the following expression for the Hamiltonian of a single vortex ring will be appropriate in our model:
			\[	H = \frac{\boldsymbol{p}^2}{2 \mu_0} +  {\beta_2{\cal E}_0} |\,b\,|^2\,.\] }
	

This interpretation of the vortex ring  in our model goes back to  Lord Kelvin's old ideas  \cite{Thom} about interpretation  of a structured particles as some  closed
 vortices.     These ideas   is still being discussed at the present time \cite{Moff}.

We will not discuss here the Hamiltonian structure of the classical theory, as well as the quantization of a single vortex. These issues are discussed in detail in the cited works of the author.

\section{The system of quantum vortices}

First of all, we need to define the Hilbert space of a quantum states.
In our  theory, the structure of this space is determined by the   independent classical variables (${\boldsymbol{p}}$,  $\boldsymbol{q}$, $\varpi$, $\chi$) which were selected above.
Indeed, the natural definition of  
 Hilbert space $\boldsymbol{H}_1$  which contains the quantum states of  a single  vortex, is following: 			
					\begin{equation}
	\label{space_quant}
	\boldsymbol{H}_1  =  \boldsymbol{H}_{pq} \otimes   \boldsymbol{H}_b\,.
	\end{equation}
			Here we have adopted the following designations. 	
			Symbol   $\boldsymbol{H}_{pq}$  denotes the Hilbert space  of a free structureless $3D$ particle  		(the space $L^2({\sf R}_3)$ {in our case }) and   symbol $\boldsymbol{H}_b $ 
						denotes the  Hilbert space of the quantum states for the  harmonic oscillator.		
						The creation and annihilation operators $\hat{b}^+$,   $\hat{b}$ as well as the standard orthonormal basis $  |\,n\rangle$
							  in the space 	$\boldsymbol{H}_b $  are defined  by well-known formulas 												
				\[ [\,\hat{b}, \hat{b}^+] = \hat{I}_b\,, \qquad \hat{b}|\,0_b\rangle = 0  \,,
				\qquad  |\,n\rangle   =  \frac{1}{\sqrt{n!}} (\hat{b}^+)^n   |\,0_b\rangle                \,.\]   
	The unit  operator in the space    $\boldsymbol{H}_b $ is  denoted as  $\hat{I}_b$.

{Our quantization  postulates for single vortex system  are  following:
					\[ q_{x,y,z} ~\to~  q_{x,y,z}\otimes \,\hat{I}_b  \,,\qquad  p_{x,y,z} ~\to~ - i\hbar\frac{\partial}{\partial q_{x,y,z}} \otimes \,\hat{I}_b \,,\]
					\[b ~\to~ \sqrt{\frac{\hbar}{t_0{\mathcal E}_0}}\, (\hat{I}_{pq} \otimes\,\hat{b})\,, \]
					where    operator $\hat{I}_{pq}$  is a unit operator in the space  $\boldsymbol{H}_{pq}$.}  					

The suggested method makes it possible  to describe the processes of creation and annihilation of closed  vortex rings of various radii.
Indeed, we can apply  the standard formalism  of  the   many-body  theory here.
Let us introduce     the  $N$-vortex space      $\boldsymbol{H}_N$:
\[ \boldsymbol{H}_N =  \bigotimes_{j=1}^N {\boldsymbol{H}}_j   \equiv  {\mathfrak H}_{pq}^N  \otimes {\mathfrak H}_b^N\,,\]

where
\[ {\mathfrak H}_{pq}^N = 
 \underbrace{\boldsymbol{H}_{pq} \otimes  \dots \otimes \boldsymbol{H}_{pq}}_{N} \,,\qquad
{\mathfrak H}_b^N =  
  \underbrace{\boldsymbol{H}_b \otimes  \dots \otimes \boldsymbol{H}_b}_{N} \,.\]
In Dirac notation, any vector  $|\Phi^N\rangle \in  \boldsymbol{H}_N$ takes the form ($N \ge 1$)
\begin{eqnarray}
\label{vect_Phi}
|\Phi^N\rangle & = & \sum_{n_1,\dots,n_N} \int\,\cdots\int d\boldsymbol{p}_1\dots\boldsymbol{p}_N 
f^N_{n_1,\dots,n_N}(\boldsymbol{p}_1,\dots,\boldsymbol{p}_N)\times \nonumber\\
~& \times & |\boldsymbol{p}_1\rangle \dots|\boldsymbol{p}_N\rangle  | n_1\rangle\dots|n_N\rangle\,,
\end{eqnarray}
where the vectors  $|\boldsymbol{p}_j\rangle$ are corresponding eigenvectors of the operators $ \hat{\boldsymbol{p}}_j$.

 The Fock  space has a structure
\[{\mathfrak H}  =  \bigoplus_{N=0}^{\infty}\boldsymbol{H}_N \subset   {\mathfrak H}_{pq} \otimes   {\mathfrak H}_{int} = {\mathfrak H}^\prime \,, \]
where   
        \[ {\mathfrak H}_{pq} = \bigoplus_{N=0}^{\infty} {\mathfrak H}_{pq}^N \,, \qquad 
{\mathfrak H}_{int} = \bigoplus_{N=0}^{\infty} {\mathfrak H}_b^N\,.  \] 
Next, we consider the ''covering''  space  ${\mathfrak H}^\prime$  only.
The creation and annihilation operators  ${\hat a}^+_{pq}(\boldsymbol{p})$,  ${\hat a}_{pq}(\boldsymbol{p})$ and  ${\hat a}^+_{int}(n)$,
${\hat a}_{int}(n)$ act in the space ${\mathfrak H}^\prime$ as ${\hat a}^+_{pq}(\boldsymbol{p})\otimes I_{int}$ and so on. 
They are defined in a standard way.
 We will also consider the operators
\[ \hat{\boldsymbol{\mathfrak a}}^+(\boldsymbol{p};n) = {\hat a}^+_{pq}(\boldsymbol{p})\otimes {\hat a}^+_{int}(n)  \,, \qquad
\hat{\boldsymbol{\mathfrak a}}(\boldsymbol{p};n) =  {\hat a}_{pq}(\boldsymbol{p})\otimes {\hat a}_{int}(n) \,\]
which act  in the  same space ${\mathfrak H}^\prime$. 

The processes of creation and annihilation of considered  vortex rings are described by  the Hamiltonian
\begin{equation}
\label{ham_q_full}
\hat{H} =   \hat{H}_0 +   \epsilon\,\hat{U}\,, 
\end{equation}
where
\begin{equation}
\label{ham_q_free}
\hat{H}_0 =   \frac{1}{2\mu_0}\int \boldsymbol{p}^2  {\hat a}^+_{pq}(\boldsymbol{p}){\hat a}_{pq}(\boldsymbol{p}) +
\frac{\beta_2 \hbar}{t_0} \sum_{n=0}^\infty \biggl(n + \frac{1}{2}  \biggr) {\hat a}^+_{int}(n){\hat a}_{int}(n)\, 
\end{equation}
and the coupling  constant $\epsilon$ is certain dimensionless constant.

The structure of  operator $\hat{H}_0$  is  motivated by group - theoretical approach to the  description  of the  energy   of a single  thin  vortex filament. 
This approach has been developed by the author in the   Ref. \cite{Tal}.
Let us note that the energy problem of a turbulent flow is still discussed in a present time (see, for example,  \cite{BrAnd}).

As example, let us consider the motion in unbounded space. 
In this case operator $\hat{H}_0$  has continuous  spectrum:
\begin{equation}
\label{spec_H0}
\hat{H}_0 \Phi^N_{\mathcal E} =  {\mathcal E} \Phi^N_{\mathcal E}\,, \nonumber
\end{equation}
where the eigenvalues ${\mathcal E}$  are positive numbers.
Let's find the eigenvectors $\Phi^N_{\mathcal E} \in \boldsymbol{H}_N $ of the operator  $\hat{H}_0$.
Introducing the designation (vector in the form of a string)
\[\Phi^N  = \bigl(0\,, 0\,,\dots, f^N_{n_1,\dots,n_N}(\boldsymbol{p}_1,\dots,\boldsymbol{p}_N)\,, 0\,,  \dots  \bigr)\,\]
and performing the direct calculations, we find for the vector $\Phi^N$:
\begin{equation}
\label{H0PhiN}
\hat{H}_0 \Phi^N \equiv  {\mathcal E}(\boldsymbol{p}_1,\dots  \boldsymbol{p}_N; n_1,\dots  n_N) \Phi^N\,,
\end{equation}
where
\begin{equation}
\label{eigen_1}
{\mathcal E}(\boldsymbol{p}_1,\dots  \boldsymbol{p}_N; n_1,\dots  n_N)  = 
  \frac{1}{2\mu_0} \sum_{j=1}^N \boldsymbol{p}_j^2   +  \frac{\beta_2 \hbar}{t_0} \sum_{j=1}^N \biggl(n_j + \frac{1}{2}  \biggr)\,.
\end{equation}
Therefore,
\begin{eqnarray}
\label{egenvect_H0}
|\Phi^N_{\mathcal E}\rangle & = & \sum_{n_1,\dots,n_N} \int\,\cdots\int d\boldsymbol{p}_1\dots\boldsymbol{p}_N 
f^N_{\mathcal E}(\boldsymbol{p}_1,\dots,\boldsymbol{p}_N; n_1,\dots  n_N)\times \nonumber\\
~& \times & |\boldsymbol{p}_1\rangle \dots|\boldsymbol{p}_N\rangle  | n_1\rangle\dots|n_N\rangle\,,
\end{eqnarray}
where the function $f^N_{\mathcal E}$ is proportional to the Dirac $\delta$-function:
\[f^N_{\mathcal E}(\boldsymbol{p}_1,\dots,\boldsymbol{p}_N; n_1,\dots  n_N) \propto \delta\Bigl( {\mathcal E} - {\mathcal E}(\boldsymbol{p}_1,\dots  \boldsymbol{p}_N; n_1,\dots  n_N) \Bigr)\,.\]
Thus, the vector $|\Phi^N_{\mathcal E}\rangle$ is the entangled state of states 
$|\boldsymbol{p}_1\rangle \dots|\boldsymbol{p}_N\rangle  | n_1\rangle\dots|n_N\rangle$
that correspond to the    $N$ 
   vortices with momenta $\boldsymbol{p}_1,\dots,\boldsymbol{p}_N$ and certain  radii ${R}_i = {R}_i(n_i) $, $i =1,\dots,N$.  

The summand  $U = U(t)$ in the formula Eq. (\ref{ham_q_full}) is the interaction hamiltonian. We discuss it in the following section in a detail.

\section{The vortex decay scenario}

So, our goal is to construct a density matrix $\hat\rho = \hat\rho(t)$ for a non-stationary and irreversible process -- the initial stage of a quantum  turbulence occurrence.
As we noted earlier, the decay of  vortices is the basis of the turbulence scenario in our model.  
Our first assumption is that we can use a non-Hermitian operator $U$ for our purposes. 
Of course, the application of non-Hermitian operators in a quantum physics has been known for a long time.  The fact that the quantum theory of irreversible processes should lead to non-unitary evolution has been discussed in book \cite{Prigogine} many years ago. Another direction is the construction of specific quantum mechanical models with non-Hermitian Hamiltonians (see, for example, \cite{Bender} and literature that cited in this paper).

The standard general representation for a density matrix for a quantum statistical system is as follows
\[\hat\rho  = \sum_k w_k |k\rangle\langle k |\,,\]
where vectors $|k\rangle$ is some pure quantum states. Therefore, we must begin our construction by defining a system of such vectors for the process under consideration.

The   suggested model   assumes   the occurrence and development of turbulence due to some single unstable vortex.
Thus, we need to define the initial  (for $t=0$) state of our system. 
We suppose that  this state   is the following pure state:
\begin{equation}
\label{init_st}
|{\boldsymbol 1}_f\rangle = \sum_n \int f_n(\boldsymbol{p}) \hat{\boldsymbol{\mathfrak a}}^+(\boldsymbol{p};n) d^3\boldsymbol{p} |0\rangle\,.
\end{equation}

We assume that function $f_n(\boldsymbol{p})$  satisfies the normalization  condition
\[ \sum_n \int |f_n(\boldsymbol{p})|^2 d^3 \boldsymbol{p} = 1\,. \]

It is clear that state $|{\boldsymbol 1}_f\rangle$ describes the single vortex with wave function  $f_n(\boldsymbol{p})$.
The index $n$ corresponds to the internal degree of the freedom here. For example,  it is possible that $f_n(\boldsymbol{p}) = f(\boldsymbol{p})\delta_{n n_0}$.
This case corresponds to the excitation of a single oscillatory level $n_0$ only.

{ As a following step,  we intend to construct the interaction Hamiltonian.
Our construction is based on the following  simplifying  assumptions.
\begin{itemize}
\item the evolution of our vortex system leads  to the  vortices decay only, but not to their unification;
\item	each vortex can decay  into two vortices  at some time moment, but no more;
\item	vortex decays occur at a random  time moments.
\end{itemize}
To implement this program, we  define the following constructions below.}

Let the value ${\sf t^v}$ means the constant that characterizes the lifetime of a vortex ({we will discuss this value  more detail later}).
Next,  we define the finite sequence of random numbers $t_k$
\begin{equation}
\label{t_random}
0 < t_1 < t_2 < \dots < t_N < t\,, \qquad  N = \left[\frac{t}{{\sf t^v}} \right] - 1\,,
\end{equation}
for every time moment $t$. 
The notation $[\,x\,]$ means the integer part of the number $x$ here. 
We  construct pure quantum states that have the following sense: the state 
\[|\Psi(t)\rangle  \equiv  | \Psi_{t_1,\dots,t_N}(t;{\sf t^v})\rangle   \] 
describes the evolution of the state
 $|{\boldsymbol 1}_f\rangle$ in the interval $[0,t]$ so that  some single vortex decays at  time moment  $t_k$, $k=1,2,\dots,N$ .

Returning to the  Eq. (\ref{ham_q_full}),
we assume that the operator $U(t)$ has following structure:
\begin{equation}
\label{U_gen}
U(t) = \sum_{n=1}^\infty U_{n\to n+1}(t)\,.
\end{equation}
{  The summand $U_{n\to n+1}(t)$ corresponds to the transition of a $n$-vortices quantum state to a state with $n+1$ vortices  at the time moment $t_n$.}

We define the operators $U_{n\to n+1}(\tau)$,  $\tau \in [0,t]$ as follows:
\begin{equation}
\label{U_n_gef}
U_{n\to n+1}(\tau) = \Bigg\{
\begin{matrix}
\hbar\delta(\tau - t_n):\Omega_1 \Omega_0^{\,n-1}:\,, & n = 1,\dots,N\,;\\[2mm]
0\,, & n > N\,.
\end{matrix}
\end{equation}
The following formulas for operators   $\Omega_0$ and   $\Omega_1$ are accepted here:
\begin{equation}
\label{Omega_0}
\Omega_0 = \sum_k \int \hat{\boldsymbol{\mathfrak a}}^+(\boldsymbol{p};k) \hat{\boldsymbol{\mathfrak a}}(\boldsymbol{p};k) d^3\boldsymbol{p}\,,
\end{equation}

\begin{eqnarray}
\label{Omega_1}
\Omega_1 & = & \sum_{m,n,k} \int \delta\left(\frac{\hbar\beta_2}{t_0}[n + m - k] - \frac{\boldsymbol{p}_1\boldsymbol{p}_2}{\mu_0}  \right)\times \\
~& \times & \hat{\boldsymbol{\mathfrak a}}^+(\boldsymbol{p}_1;n) \hat{\boldsymbol{\mathfrak a}}^+(\boldsymbol{p}_2;m)
\hat{\boldsymbol{\mathfrak a}}(\boldsymbol{p}_1 + \boldsymbol{p}_2 ;k) d^3\boldsymbol{p}_1 d^3\boldsymbol{p}_2\,. \nonumber
\end{eqnarray}
Symbol $:~~:$ means the normal ordering. 
{ Regarding the coefficient functions $K(\dots)$ in the integrals, those functions are motivated by the conservation of the energy.  So, the function $K$ is identical in the  Eq. (\ref{Omega_0}) and $K = \delta[({\hbar\beta_2}/{t_0})[n + m - k] - {\boldsymbol{p}_1\boldsymbol{p}_2}/{\mu_0}]$ in the  Eq. (\ref{Omega_1}). }
 Let us note that
\[U_{n\to n+1}(\tau)U_{m\to m+1}(\tau) = 0\,,\qquad  n \not= m\,.\]
The operators  $:\Omega_1 \Omega_0^{\,n-1}:$  have the following important property. Let the vector $|\Psi_m\rangle \in {\boldsymbol H}_m$ is any vector which describes 
some $m$ - vortex state. For example,
\[|\Psi_m\rangle = \prod_{j=1}^m \hat{\boldsymbol{\mathfrak a}}^+(\boldsymbol{p}_j;n_j) |0\rangle\,.\]
Then 
\[ :\Omega_1 \Omega_0^{\,n-1}:|\Psi_m\rangle = 0\,, \qquad m = 1,\dots, n-1\,.\]

We  consider the evolution of vectors $|\Psi\rangle \in {\mathfrak H}$ in the interaction  representation. 
Therefore, returning our consideration on the time interval $[0,t]$, we can write
\[ V(\tau) = e^{\frac{i}{\hbar}H_0 \tau}U(\tau)e^{-\frac{i}{\hbar}H_0 \tau} = \hbar\sum_{n=1}^N \delta(\tau - t_n) V_{n}\,, \quad  \tau \in  [0,t]\,, \]
where operator $H_0$ is defined by the  Eq. (\ref{ham_q_free}) and operators $V_{n}$ are defined as follows:
\begin{equation}
\label{V_n}
  V_{n} = e^{\frac{i}{\hbar}H_0 t_n}\,:\Omega_1 \Omega_0^{\,n-1}:\,e^{-\frac{i}{\hbar}H_0 t_n}\,.
\end{equation}
We  describe the evolution of the vectors $|\Psi\rangle \in {\mathfrak H}$  in interaction representation in a standard way:
\begin{equation}
\label{evol_gen}
|\Psi(t^{\prime\prime})\rangle = {\cal U}(t^{\prime\prime},t^{\prime})|\Psi(t^{\prime})\rangle \,,\qquad  t^{\prime\prime} > t^\prime \,,
\end{equation}
where symbol  ${\cal U}(t^{\prime\prime},t^{\prime})$ means the Dyson series that has to  constructed with help of the operator $V(t)$.

The formula Eq. (\ref{evol_gen}) should be factorized on the interval $[0,t]$ for our purposes.
In order to   { perform this procedure}, we consider arbitrary numbers $\xi_i$, $i = 1,\dots, N-1$  that satisfy the conditions
\begin{equation}
\label{xi_random}
0 < t_1 < \xi_1 < t_2 <\xi_2 < \dots < \xi_{N-1} < t_N < t \,.
\end{equation}
Then we  represent the evolution operator in the following form:
\begin{equation}
\label{evol_factor}
|\Psi(t)\rangle = {\cal U}(t, \xi_{N-1}) {\cal U}(\xi_{N-1}, \xi_{N-2})\dots {\cal U}(\xi_1, 0) |\Psi(0)\rangle \,,
\end{equation}
Assuming that the coupling constant $\epsilon$ (see   Eq. (\ref{ham_q_full})) is small enough, we  use  first Born approximation for the operators
${\cal U}(\xi_{n}, \xi_{n-1})$.
 Thus, we have the following representation for the vector $|\Psi(t)\rangle$:
\begin{equation}
\label{Psi_evol}
|\Psi(t)\rangle = \Bigl(1 - i\epsilon V_N \Bigr)\Bigl(1 - i\epsilon V_{N-1} \Bigr)\dots    \Bigl(1 - i\epsilon V_1 \Bigr)|{\boldsymbol 1}_f\rangle\,.
\end{equation}
It is clear that the representation Eq. (\ref{Psi_evol}) does not depend on the choice of numbers  $\xi_i$.

\begin{figure}[h]
\includegraphics[width=3.37in]{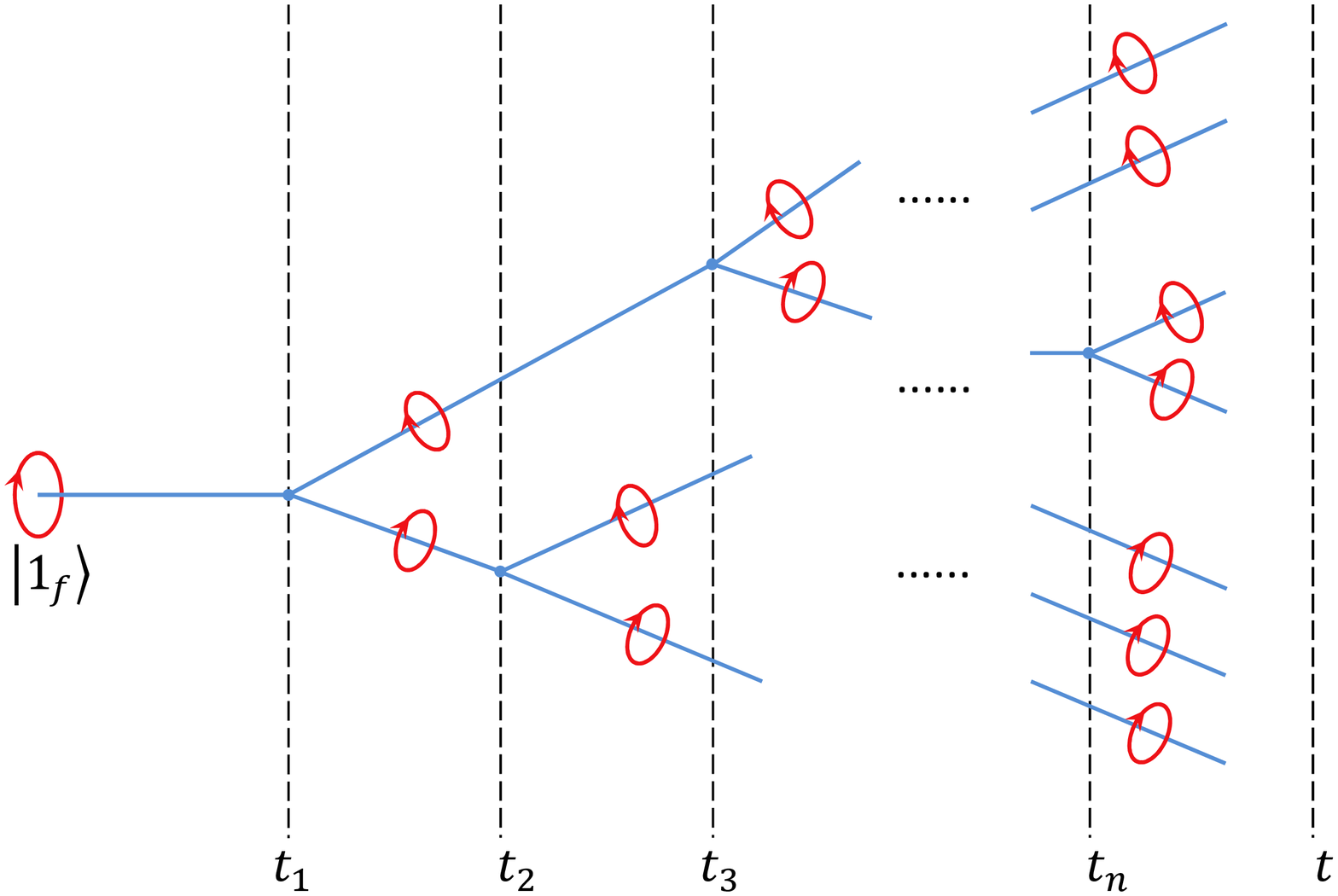}
\caption{vortices decay scenario}
\label{Vort_1}
\end{figure}

Let's demonstrate how vector $|\Psi(t)\rangle$ evolves according to the  Eq. (\ref{Psi_evol}).
Indeed, let us suppose that value  $t$ be small enough so that $N=1$. It is clear that 
\[ |\Psi(t)\rangle  \equiv  | \Psi_{t_1}(t;{\sf t^v})\rangle   = \Bigl(1 - i\epsilon V_1 \Bigr)|{\boldsymbol 1}_f\rangle \]
in this case.
By virtue of the definition Eq. (\ref{V_n}) operator $V_1$ has following form:
\[V_{1} = e^{\frac{i}{\hbar}H_0 t_1}\,\Omega_1\,e^{-\frac{i}{\hbar}H_0 t_1}\,.\]
Operators  $e^{\pm\frac{i}{\hbar}H_0 t_1}$  map any space $\boldsymbol{H}_N$ into themselves; operator $\Omega_1$ implements the mapping
$\boldsymbol{H}_1 \to \boldsymbol{H}_2$.  Consequently,
\[ |\Psi(t)\rangle  \equiv  | \Psi_{t_1}(t;{\sf t^v})\rangle \in \boldsymbol{H}_1 \oplus \boldsymbol{H}_2\,.\]
Similarly, we find that  the following inclusion is true  in  general case:
\[|\Psi(t)\rangle  \equiv  | \Psi_{t_1,\dots,t_N}(t;{\sf t^v})\rangle \in \boldsymbol{H}_1 \oplus\dots\oplus \boldsymbol{H}_N\,.  \] 
Finally, we constructed the pure quantum state which corresponds to the  evolution of initial vortex in accordance with certain scenario.
This scenario suppose the    decay  { of}  one  { of}  the  vortex rings  at fixed  time moment $t_n$, where $n = 1,\dots,N$. 
The first Born approximation means that we take into account  the decay of the vortex into two vortices only.  {  This scenario is clearly illustrated by the figure \ref{Vort_1}.}

Given the subsequent terms of the Dyson  series for the operator ${\cal U}(t^{\prime\prime},t^{\prime})$, 
we can include in the consideration { more complicated events.}
However, we will not consider these       processes here.

\section{Density matrix}

In this section we  construct the density matrix  $\hat\rho$  in the final form. 
Taking into account our previous considerations, we can write the formula
\begin{eqnarray}
\label{rho_1}
\hat\rho(t) & = & \frac{1}{(\Delta{\sf t})^{N+1}}\int_{0}^{\Delta{\sf t}} d{\sf t^v}
\int_0^t \dots \int_0^t dt_1\dots dt_N \,\times \nonumber \\[2mm]
~~ & \times & w({\sf t^v}; t_1,\dots,t_N)\, | \Psi_{t_1\dots t_N}(t;{\sf t^v})\rangle \langle \Psi_{t_1\dots t_N}(t;{\sf t^v})|\,,
\end{eqnarray}
where  the function  $w({\sf t^v}; t_1,\dots,t_N)$ means the probability of an  initial state    evolution  in accordance with scenario  which was described earlier.

To define the function $w({\sf t^v}; t_1,\dots,t_N)$, we  make the following assumptions:

\begin{em}
\begin{itemize}
 \item The probability { of the event} that a single vortex  still  exists during the time interval $[0,t]$ is determined by the formula
\begin{equation}
\label{p1}
p_1({\sf t^v}; t) = \exp\Bigl(-\frac{t}{{\sf t^v}} \Bigr)\,.
\end{equation}
\item Only one vortex can decay at any time moment $t$.
\end{itemize}
\end{em} 

Correspondingly, the probability that a single vortex   decays  
at the time moment $t$    is determined by the formula
\begin{equation}
\label{w1}
w_1({\sf t^v}; t) =1 - \exp\Bigl(-\frac{t}{{\sf t^v}} \Bigr)\,.
\end{equation}

We believe that the vortices resulting from the  series   of the  vortex decays are independent objects.
Let us suppose that we have $k$ vortices in the time interval $[t_{k-1}, t_k]$ where $k = 1,\dots,N$ and $t_0 = 0$.
Then, the probability that  the  only  vortex  decays at the moment $t = t_k$ is determined by the formula:

\begin{equation}
\label{wk}
w_k({\sf t^v}; t_{k-1}, t_k) =\theta(t_k - t_{k-1})\left[1 - \exp\Bigl(-\frac{ t_k - t_{k-1}}{{\sf t^v}} \Bigr)\right] \exp\Bigl(-(k-1)\frac{ t_k - t_{k-1}}{{\sf t^v}} \Bigr) \,.
\end{equation}

Therefore, the following formula for the function $w({\sf t^v}; t_1,\dots,t_N)$ is appropriate:
\begin{equation}
\label{w_final}
 w({\sf t^v}; t_1,\dots,t_N)   =  \prod_{k=0}^N w_k({\sf t^v}; t_{k-1}, t_k)  \,.
\end{equation}
Let us note that Heaviside functions  $\theta(t_k - t_{k-1})$ provide the fulfillment of the condition Eq. (\ref{t_random}) when integrating in the formula Eq. (\ref{rho_1}).

To complete the construction of the density matrix, we must justify the constant  $\Delta{\sf t}$ as the upper limit in the  Eq. (\ref{rho_1}).

Following the Ref. \cite{Tenn}, we  assume that
\[  {\sf t^v} = T\,,\]
where constant $T$ means the period of the vortex rotation.
Let the values $\omega$, ${\sf a}$ and $v_a$ mean the rotation frequency, vortex core radius and the fluid velocity on the core boundary.
Then the following simple formulas are valid:
\[  T = \frac{2\pi}{\omega}\,, \qquad 
\Gamma= 2\pi {\sf a} v_a = 2\pi {\sf a}^2\omega\,.\] 
Therefore,
\[  {\sf t^v} = \frac{4\pi^2 {\sf a}^2}{\Gamma}\,.\]

We consider the vortices with a small (${\sf a} \simeq 0$) core radius only.
The value ${\sf a}$ is connected with the vortex ring velocity $\boldsymbol{u}_{v}$ (See Ref.\,\cite{Saffm}):
\[ |\boldsymbol{u}_{v}|  \simeq  \frac{\Gamma}{4\pi R}\ln\frac{8R}{{\sf a}}\,.\]

Because we consider the solutions Eq. (\ref{our_sol}) only, the value  $|\boldsymbol{u}_{v}| = \beta_1 R/t_0$ in our model.
Thus, the following expression   is true  for the value ${\sf a}$:
\begin{equation}
        \label{core_radius}
		{\sf a} \simeq 8 R \exp\left(-  \frac{4\pi\beta_1 R^2}{t_0 \Gamma}  \right)\,.		
\end{equation}
Using these formulas, we find the function that determines the lifetime ${\sf t^v}$ of the vortices under consideration:
\begin{equation}
        \label{tau_F}
			{\sf t^v} = F(\eta) \equiv 32\pi \eta \exp\Bigl(-\frac{\beta_1\eta}{t_0} \Bigr)\,,	\qquad  \eta = \frac{8\pi R^2}{\Gamma}\,.\nonumber
	\end{equation}			

This function has maximum value 
\begin{equation}
        \label{tau_max}
		{\sf t^v}_{max} = \frac{32\pi t_0}{\beta_1 e}		
\end{equation}
at the point $\eta = t_0/\beta_1$.

On the other hand,  the uncertainty   in determining the energy for the initial state under consideration Eq. (\ref{init_st}) is the value (see, for example,  Ref. \cite{LanLif})
\[  \delta{\cal E}_f = \sqrt{\langle{\boldsymbol 1}_f|  \hat{H}_0^2 |{\boldsymbol 1}_f\rangle  -  \langle{\boldsymbol 1}_f|  \hat{H}_0 |{\boldsymbol 1}_f\rangle^2 }\,.\]
Taking into account the ''energy-time''  Heisenberg uncertainty ratio, we can assume that   uncertainty   in determining of time is given  by the formula
\[ \delta t \simeq \frac{\hbar}{\delta{\cal E}_f}\,.\]
 Therefore, the following determination of the constant  $\Delta{\sf t}$ is appropriate:
\begin{equation}
        \label{Delta_t}
			\Delta{\sf t} = \Biggl\{
				\begin{matrix}
				~~{\sf t^v}_{max}\,, & \qquad \delta t  < {\sf t^v}_{max}\,,\\[2mm]
				\frac{\hbar}{\delta{\cal E}_f}\,, & \qquad \delta t  \ge {\sf t^v}_{max}\,.
				\end{matrix}
	\end{equation}

\section{Concluding remarks}
The model proposed in this paper demonstrates the scenario of the evolution of a single quantum vortex ring into a system of $N$ quantum vortex rings.
{ We interpret this process as the initial stage of turbulence development.
 The suggested approach provides a fundamental opportunity to calculate the partition function
\begin{equation}
  {\mathcal Z}  = {\rm Tr} \exp\left(- \frac{\hat H}{{\sf k}_B{\sf T}}\right)\,,\nonumber
\end{equation}
where  value ${\sf T}$ is the  temperature and the value ${\sf k}_B$ is  Boltzmann constant.
The spectrum of the Hamiltonian $\hat H$ can be either continuous or discrete, depending on the boundary conditions. This possibility  is important for the subsequent study of the thermodynamic characteristics of the turbulent flow.  Moreover, the proposed model leads to a circulation spectrum that is more complex than the standard spectrum $\Gamma_n \propto n$, where the value $n$ is integer number. From our point of view, the spectrum of fluid circulation in a turbulent flow may be more complex. Let us note that quantization of circulation is one of the main points that distinguish quantum turbulence from classical ones.}

Of course, a realistic scenario also involves some series of moments that are not taken into account here. First, we must take into account 
changing the shape of the rings. Such effects can be investigated with help of the perturbation theory using the methods developed in Ref. \cite{Tal}.
Second, the vortex interaction suppose both decay and unification processes.  These processes require additional terms in the interaction Hamiltonian.
More complicated problem is consideration of tangled and knotted vortices (See Ref.\cite{Baren} for example).  The author hopes to return to these issues in subsequent works.

\end{document}